\documentclass[12pt,oneside]{article}
\usepackage{epsfig, cite}
\usepackage{color}
\usepackage{ifthen}
\usepackage{graphicx}

\def\half{{1\over 2}}
\def\({\left(}
\def\){\right)}
\def\[{\left[}
\def\]{\right]}

\def\e{\begin{equation}}
\def\q{\end{equation}}
\def\m{\begin{eqnarray}}
\def\n{\end{eqnarray}}

\begin{document}
\thispagestyle{empty} \setcounter{page}{0}

\vspace{2cm}

\begin{center}
{\huge Spectral Index in Curvaton Scenario}

\vspace{1.4cm}

Qing-Guo Huang

\vspace{.2cm}

{\em School of physics, Korea Institute for
Advanced Study,} \\
{\em 207-43, Cheongryangri-Dong,
Dongdaemun-Gu, } \\
{\em Seoul 130-722, Korea}\\
\end{center}

\vspace{-.1cm}

\centerline{{\tt huangqg@kias.re.kr}} \vspace{1cm}
\centerline{ABSTRACT}
\begin{quote}
\vspace{.5cm}

A red tilted primordial power spectrum is preferred by WMAP
five-year data and a large positive local-type non-Gaussianity
$f_{NL}$ might be observed as well. In this short note we find that
a red tilted and large non-Gaussian primordial power spectrum cannot
be naturally obtained in curvaton model, because $f_{NL}$ is related
to the initial condition of inflation.

\end{quote}
\baselineskip18pt

\noindent

\vspace{5mm}

\newpage


Inflation \cite{Guth:1980zm} provides an elegant mechanism to solve
many puzzles in the Hot Big Bang model. The wrinkles in the cosmic
microwave background radiation and the large-scale structure of the
Universe are seeded by the quantum fluctuations generated during
inflation \cite{Guth:1982ec}. The shape of the primordial quantum
fluctuations is characterized by its amplitude $P_\zeta$ and tilt
$n_s$ which can be measured by experiments. WMAP five-year data
\cite{Komatsu:2008hk} combined with the distance measurements from
the Type Ia supernovae (SN) and the Baryon Acoustic Oscillations
(BAO) in the distribution of galaxies indicates \m
P_{\zeta,obs}&=&2.457_{-0.093}^{+0.092}\times 10^{-9}, \\
n_s&=&0.960_{-0.013}^{+0.014}. \n Gravitational wave perturbations
are also generated during inflation and its amplitude $P_T$ is only
determined by the inflation scale. For convenience, we define a new
quantity, named tensor-scalar ratio $r=P_T/P_\zeta$, to measure the
amplitude of gravitational wave perturbations. The primordial
gravitational wave perturbation has not been detected. Present limit
on the tensor-scalar ratio is $r<0.20$ (95$\%$ \hbox{CL}). The blue
tilted primordial power spectrum $(n_s>1)$ is disfavored even when
gravitational waves are included.

The non-Gaussianity is characterized by the non-Gaussianity
parameter $f_{NL}$ which is defined as follows \e \Phi({\bf
x})=\Phi_L({\bf x})+f_{NL}[\Phi_L^2({\bf x})-\langle\Phi_L^2({\bf
x})\rangle], \q where $\Phi_L({\bf x})$ denotes the linear Gaussian
part of the perturbation in real space. The simplest model of
inflation predicts a closely Gaussian distribution of primordial
fluctuations \cite{Maldacena:2002vr,Bartolo:2004if}, namely
$|f_{NL}|<{\cal O}(1)$. The most general density perturbation is a
superposition of an isocurvature density perturbation and an
adiabatic density perturbation. We introduce a new parameter
$\alpha_{-1}=f_{iso}^2/(1+f_{iso}^2)$ to measure the isocurvature
density perturbation, where $f_{iso}$ is the ratio of the
isocurvature and adiabatic amplitudes at the pivot scale. A Gaussian
and adiabatic power spectrum of primordial perturbation is still
consistent with WMAP five-year data: \e -9<f_{NL}^{local}<111  \quad
\hbox{and} \quad -151<f_{NL}^{equil}<253 \quad (95\% \hbox{CL}), \q
\e \alpha_{-1}<0.0037 \quad (95\% \hbox{CL}), \q where ``local" and
``equil" denote the shapes of the non-Gaussianity. In
\cite{Yadav:2007yy} the authors reported that a positive large
non-Gaussianity \e 27<f_{NL}^{local}<147 \q is detected at $95\%$
C.L.. A large non-Gaussianity is not a conclusive result from
experiments, but it is still worthy for us to discussing the
theoretical probabilities of the large non-Gaussianity. If it is
confirmed by the forthcoming cosmological experiments, it strongly
shows up some very important new physics of the early Universe.

An attractive model for a large positive local-type non-Gaussianity
is curvaton model \cite{Enqvist:2001zp,Lyth:2001nq} in which the
primordial power spectrum is generated by a light scalar field,
called curvaton $\sigma$, but not the inflaton $\phi$, even though
the dynamics of inflation is governed by the inflaton. Recently many
issues about curvaton model were discussed in
\cite{Huang:2008ze,Ichikawa:2008iq,Multamaki:2008yv,Beltran:2008ei}.
In \cite{Huang:2008ze} we considered the case in which the Hubble
parameter is roughly a constant during inflation and found that
$f_{NL}$ is bounded by the tensor-scalar ratio $r$ from above. Or
equivalently a large non-Gaussianity gives a lower bound on the
amplitude of the tensor perturbation in curvaton scenario. In
\cite{Komatsu:2008hk} the authors pointed out that a large positive
$f_{NL}$ cannot be obtained by considering the bound on the
isocurvature perturbation $\alpha_{-1}<0.0037$ in curvaton model.
However if the cold dark matter was produced after the curvaton
decays completely, the curvaton model is free from the constraint of
isocurvature perturbation \cite{Beltran:2008ei}. Some other related
topics on the large non-Gaussianity are discussed in
\cite{Chen:2006nt} recently.

The spectral index of primordial power spectrum in curvaton scenario
is given by \e n_s=1-2\epsilon+2\eta_{\sigma\sigma}, \label{index}\q
where $\eta_{\sigma\sigma}={1\over 3H^2}{d^2V(\sigma)\over
d\sigma^2}$ and \e \epsilon\equiv -{\dot H/H^2}\q denotes how fast
the Hubble parameter varies during inflation. Usually in curvaton
model we assume that the mass of curvaton ${d^2V(\sigma)\over
d\sigma^2}$ is much smaller than the Hubble parameter during
inflation. Therefore $n_s\simeq 1-2\epsilon$. For $n_s=0.96$,
$\epsilon\simeq 0.02$. Such a large value of $\epsilon$ implies the
variation of inflaton is larger than Planck scale, which might be
inconsistent with string theory
\cite{Huang:2007gk,Chen:2006hs,Ooguri:2006in,McAllister:2007bg}.
Usually a small value of $\epsilon$ and a closely scale-invariant
power spectrum are expected in curvaton scenario. However, maybe the
effective tran-Planckian excursion of inflation can be achieved in
string landscape \cite{Huang:2007ek} or the monodromies
\cite{Silverstein:2008sg}.

In this short note we will extend our previous work
\cite{Huang:2008ze} to more general cases and investigate whether a
red tilted primordial power spectrum is naturally compatible with a
large positive $f_{NL}$ in curvaton model.

Let us consider a simple curvaton model with potential \e
V(\phi,\sigma)=V(\phi)+\half m^2\sigma^2, \q where $V(\phi)\gg \half
m^2\sigma^2$, $\phi$ and $\sigma$ denote the inflaton and curvaton
respectively. The slow-roll equations of motion are obtained by
neglecting the kinetic term and the energy density of curvaton in
the Friedmann equation, and the second time derivative in the
inflaton field equation, namely \m H^2&=&{V(\phi)\over 3M_p^2}, \\
3H\dot \phi&=&-V'(\phi), \n where $V'(\phi)=dV(\phi)/d\phi$ and
$M_p$ is the reduced Planck scale. Once inflation is over, the
energy density of inflaton is converted into radiation and $H^2$
goes like $a^{-4}$. The value of curvaton field, denoted as
$\sigma_*$, does not change until the Hubble parameter becomes the
same order of curvaton mass. After that curvaton oscillates around
its minimum $\sigma=0$ and its energy density decreases as $a^{-3}$.
Once the Hubble parameter goes to the same order of the curvaton
decay rate $\Gamma_\sigma$, the curvaton energy is converted into
radiations. Before primordial nucleosynthesis, the curvaton field is
supposed to completely decay into radiation and thus the
perturbations in the curvaton field are converted into curvature
perturbations. The amplitude of the perturbations caused by curvaton
is given in \cite{Lyth:2001nq} by \e P_{\zeta_\sigma}^\half={1\over
3\pi}\Omega_{\sigma,D}{H_*\over \sigma_*}, \label{ps}\q where \e
\Omega_{\sigma,D}\equiv\({\rho_\sigma\over \rho_{tot}}\)_D \q is the
fraction of curvaton energy density in the energy budget at the time
of $H=\Gamma_\sigma$. Here the subscript * denotes that the
quantities are evaluated at horizon exit during inflation. A large
positive non-Gaussianity is obtained only when $\Omega_{\sigma,D}\ll
1$ and $f_{NL}$ is given by \e f_{NL}={2\over 3}-{5\over
6}\Omega_{\sigma,D}+{5\over 4\Omega_{\sigma,D}}\simeq {5\over
4\Omega_{\sigma,D}}. \label{fnm} \q In the case of
$\Omega_{\sigma,D}\ll 1$ the Universe is dominated by radiation
before the time of $H=\Gamma_\sigma$. In \cite{Lyth:2001nq} the
value of $\Omega_{\sigma,D}$ is estimated as \e
\Omega_{\sigma,D}\simeq {\sigma_*^2\over 6M_p^2}\({m\over
\Gamma_\sigma}\)^\half.\label{ggm}\q Keeping $m$ and $\Gamma_\sigma$
fixed, $f_{NL}\propto M_p^2/\sigma_*^2$. In the literatures
$\sigma_*$ is treated as a free parameter and then a large $f_{NL}$
is naturally expected for $\sigma_*<M_p$. However this treatment
seems too naive. The curvaton mass is much smaller than the Hubble
parameter during inflation, which means the Compton wavelength is
large compared to the curvature radius of the de Sitter space
$H^{-1}$. So the gravitational effects may play a crucial role on
the behavior of the light scalar field in such a scenario. In
\cite{Bunch:1978yq} the authors explicitly showed that the quantum
fluctuation of the light scalar field $\sigma$ with mass $m$ in de
Sitter space gives it a non-zero expectation value of the square of
a light scalar field \e \langle\sigma^2\rangle={3H^4\over
8\pi^2m^2},\label{bdw} \q where the Hubble parameter $H$ is assumed
to be a constant. In \cite{Huang:2008ze} we estimated the value of
curvaton as $\sigma_*\sim H^2/m$ and we found $f_{NL}<522r^{1\over
4}$.

It is more complicated to estimate the value of curvaton in the
models, such as chaotic inflation, where the Hubble constant cannot
be regarded as a constant. Fortunately, how to estimate the value of
curvaton field in this case has been discussed by Linde and Mukhanov
in \cite{Linde:2005yw}. According to the long-wave quantum
fluctuation of a light scalar field $(m\ll H)$ in inflationary
universe, its behavior can be taken as a random walk
\cite{Linde:2005ht}: \e \langle\sigma^2\rangle={H^3\over 4\pi^2}t.
\label{ss} \q This equation is only valid for the case of a constant
$H$. During inflation the Hubble parameter $H$ is not a constant
exactly. We don't explicitly know the behavior of
$\langle\sigma^2\rangle$ when $H$ depends on $t$. However, for a
short period $\Delta t$ $(\ll H^{-1})$, Eq.(\ref{ss}) can be written
as \e \Delta \langle\sigma^2\rangle\simeq {H^3\over
4\pi^2}(1-3\epsilon H\Delta t)\Delta t. \q Since $3\epsilon H\Delta
t\ll 1$, we suppose that the differential form of Eq.(\ref{ss}) \e
{d\langle\sigma^2\rangle\over dt}\simeq {H^3\over 4\pi^2}\q can be
generalized to the case in which the Hubble parameter slowly varies
$(\epsilon=-\dot H/H^2\ll 1)$. On the other hand, a massive scalar
field cannot grow up to arbitrary large vacuum expectation value
because it has a potential. The long wavelength modes of the light
scalar field are in the slow-roll regime and obey the slow-roll
equation of motion, i.e.  \e 3H{d\sigma\over dt}=-{dV(\sigma)\over
d\sigma}=-m^2\sigma.\q Combining these two considerations, Linde and
Mukhanov proposed in \cite{Linde:2005yw} \e
{d\langle\sigma^2\rangle\over dt}={H^3\over 4\pi^2}-{2m^2\over
3H}\langle\sigma^2\rangle. \label{rw}\q For a constant Hubble
parameter, $\langle\sigma^2\rangle$ stabilizes at the point of $
\langle\sigma^2\rangle={3H^4\over 8\pi^2m^2}$ which is just the same
as Eq.(\ref{bdw}). Classically the scalar field is stable at
$\sigma=0$. In the inflationary universe the scalar field $\sigma$
gets a non-zero expectation value due to the gravitational effects.
Integrating over Eq.(\ref{rw}) with the initial condition
$\langle\sigma^2(t=t_i)\rangle=0$, we obtain \e
\langle\sigma^2(t)\rangle=\int_{t_i}^{t}dt_1{H^3(t_1)\over
4\pi^2}\exp\(-\int_{t_1}^tdt_2{2m^2\over 3H(t_2)}\). \q We will use
this solution to estimate the value of curvaton field.

Let's take into account the chaotic inflation which is driven by the
potential \e V(\phi)={1 \over p}\lambda \phi^p M_p^{4-p},
\label{pt}\q where $\lambda$ is a small dimensionless parameter
($\lambda\ll 1$) and $p>0$. From now on we work in the unit of
$M_p=1$. The equations of motion for the slow-roll inflation are
given by \m H^2&=&{\lambda \phi^p\over 3p}, \label{srh}
\\ 3H\dot \phi&=&-\lambda \phi^{p-1}. \label{srp}\n The value of inflaton
at the time of $N$ e-folds before the end of inflation is related to
$N$ by \e \phi_N=\sqrt{2p N}.\q Now the slow-roll parameter
$\epsilon$ becomes \e \epsilon=-{\dot H\over H^2}={1\over
2}\({V'\over V}\)^2={p^2\over 2\phi_N^2}={p\over 4N}. \q The number
of e-folds corresponds to the CMB scale is roughly $N_c=60$. For
$n_s=0.96$, $p\simeq 4.8$. The amplitude of scalar primordial power
spectrum caused by inflaton at CMB scale is \e P_\phi={H^2\over
8\pi^2\epsilon}={1\over 12\pi^2 p^3}\lambda\phi_{N_c}^{p+2}. \q In
curvaton scenario $P_\phi\leq P_{\zeta,obs}$ which induces an upper
bound on $\lambda$, namely \e \lambda\leq
12\pi^2p^3P_{\zeta,obs}\phi_{N_c}^{-(p+2)}. \label{bl}\q For $p=2$
the mass of inflaton $\sqrt{\lambda}$ should be smaller than
$6.36\times 10^{-6}$ in unit of Planck scale. On the other hand, the
curvaton mass is smaller than Hubble parameter which says \e m^2\leq
{\lambda \phi^p \over 3p}\simeq {\lambda \over 3p}. \q In the last
step we consider that $\phi\sim M_p$ at the end of inflation.
Combing with Eq.(\ref{bl}), we have \e m^2\leq
4\pi^2p^2P_{\zeta,obs}\phi_{N_c}^{-(p+2)}. \label{mh}\q For $p=2$,
the curvaton mass satisfies $m\leq 2.6\times 10^{-6}$.

It is time to estimate the vacuum expectation value of curvaton in
chaotic inflationary universe. Using the equations of motion for the
slow-roll chaotic inflation (\ref{srh}) and (\ref{srp}), we simplify
Eq.(\ref{rw}) to be \e \langle\sigma^2(t)\rangle\simeq {\lambda\over
12\pi^2p^2}\int_{\phi(t)}^{\phi_i}d\phi_1\phi_1^{p+1}\exp\({2m^2\over
\lambda}\int_{\phi_1}^\phi d\phi_2\phi_2^{-(p-1)}\)\simeq {\lambda
\over 12\pi^2p^2(p+2)}\phi_i^{p+2}, \q where we ignore the
contribution from the exponential function because its exponent is
proportional to $m^2/m_{eff}^2(\phi)\sim m^2/(\lambda\phi^{p-2})\ll
1$. This vacuum expectation value of curvaton mainly comes from the
perturbation mode with wavelength $H^{-1}(\phi_i)\exp({\phi_i^2\over
2p})$. Since the wavelength is much larger than the Hubble horizon,
this fluctuation mode is frozen to be a classical one and provides a
non-zero classical configuration for curvaton field. The typical
value of curvaton field in such a background is estimated as \e
\sigma_*^2={\lambda \over 12\pi^2p^2(p+2)}\phi_i^{p+2}, \label{sg}
\q which  is obviously dependent on the initial value of inflaton.
Requiring the curvaton energy density be much smaller than inflaton
energy density during inflation yields a upper bound on the curvaton
mass \e m^2\ll 24\pi^2p(p+2)\phi_i^{-(p+2)}. \label{mg} \q Here we
also ignore a factor $\phi^p$ on the right hand side of the above
inequality because $\phi\sim 1$ at the end of inflation. Since the
total number of e-folds of chaotic inflation is not very large, the
constraint in Eq.(\ref{mh}) is much more stringent than
Eq.(\ref{mg}).

In curvaton scenario the primordial power spectrum comes from the
quantum fluctuation of curvaton during inflation. Substituting
$\sigma_*$ in Eq.(\ref{sg}) into (\ref{ps}), we get \e
P_{\zeta_\sigma}={4p(p+2)\over 9}\Omega_{\sigma,D}^2{\phi_{N}^p\over
\phi_i^{p+2}}. \label{pzs}\q Now the spectral index of primordial
power spectrum is given by \e n_s\equiv 1+{d\ln
P_{\zeta_\sigma}\over d\ln k}\simeq 1-{d\ln P_{\zeta_\sigma}\over
dN}=1-{p\over 2N}.\q Using Eq.(\ref{fnm}), we write down $f_{NL}$ as
follows \e f_{NL}={5\sqrt{p(p+2)}\over
6}P_{\zeta,obs}^{-1/2}{\phi_{N}^{p/2}\over \phi_i^{p/2+1}}.
\label{fin}\q We need to stress that $f_{NL}$ depends on the initial
condition of inflation or the total number of e-folds
$N_t=\phi_i^2/(2p)$! For $p=2$, $f_{NL}=1.84\times 10^5/N_t$. If
$f_{NL}=100$, $N_t=1.84\times 10^3$. The problem is why inflation
has such an initial condition. It is very hard to give a physical
explanation on it. In this sense, curvaton model can not naturally
explain a large non-Gaussian and red tilted primordial power
spectrum.

Is it possible that eternal chaotic inflation \cite{Linde:1986fc}
offers a proper initial condition? During the period of inflation,
the evolution of the inflaton field $\phi$ is also influenced by
quantum fluctuations, which can also be pictured as a random walk of
inflaton with a step $\delta \phi={H\over 2\pi}$ on a horizon scale
per Hubble time $H^{-1}$. During the same epoch, the variation of
the classical homogeneous inflaton field rolling down its potential
is $\Delta \phi=|\dot \phi|\cdot H^{-1}$. Eternal chaotic inflation
happens when $\delta \phi=\Delta \phi$, namely \e
\phi=\phi_E=\(12\pi^2p^3/\lambda\)^{1\over p+2}\q for the chaotic
inflation with potential Eq.(\ref{pt}). Naively we take $\phi_E$ as
the initial value of inflaton $\phi_i$. Therefore we have \e
\sigma_*^2={p\over p+2}, \q and the primordial power spectrum
generated by curvaton \e P_{\zeta_\sigma}={p+2\over
27\pi^2p^2}\lambda\phi_{N_c}^p\Omega_{\sigma,D}^2. \q Considering
that $\lambda$ is bounded from above by (\ref{bl}) and
$P_{\zeta_\sigma}=P_{\zeta,obs}$, we find \e \epsilon\geq {9\over
8}\Omega_{\sigma,D}^{-2}-{1\over 2N_c}. \q The slow-roll parameter
$\epsilon$ must be larger than one because $\Omega_{\sigma,D}\leq 1$
and the slow-roll condition is violated. The reason is that the
energy scale of eternal chaotic inflation is quite high and curvaton
gets a large vacuum expectation value which strongly suppresses the
amplitude of primordial power spectrum to be smaller than
$P_{\zeta,obs}$. So we conclude that inflation should start at an
energy scale lower than eternal inflation scale for the validity of
curvaton model.

Before the end of this note, we ignore the initial condition problem
and investigate the possible parameter space for curvaton model with
a large non-Gaussianity. Following \cite{Lyth:2003dt}, the decay
rate of curvaton $\Gamma_\sigma$ should be greater than the
gravitional-strength decay rate $m^3$ in the unit of Planck scale.
Combining with Eq.(\ref{fnm}) and (\ref{ggm}), we obtain \e m\leq
{2\over 15}f_{NL}\sigma_*^2. \q Using Eq.(\ref{bl}), (\ref{sg}) and
(\ref{fin}), after a straightforward calculation we find \e
\sigma_*^2\leq {25p\over 72N_c}f_{NL}^{-2}. \q Combining the above
two inequalities leads to an upper bound on the curvaton mass \e
m\leq 7.7\cdot 10^{-4}\cdot p/f_{NL}. \q If $f_{NL}\sim 100$, this
constraint is roughly the same order as Eq.(\ref{mh}). In order to
obtain a large non-Gaussianity the curvaton mass should be smaller
than $10^{12}$ GeV in the curvaton model combined with chaotic
inflation. On the other hand, the curvaton should decay before
neutrino decoupling \cite{Lyth:2003dt}; otherwise the curvature
perturbations may be accompanied by a significant isocurvature
neutrino perturbation. The temperature of the universe at the moment
of neutrino decoupling is roughly $T_{nd}=1$ MeV. The curvaton decay
rate is bounded by the Hubble parameter at the time of neutrino
decoupling from below, i.e. $\Gamma_\sigma>\Gamma_0=1.68\times
10^{-43}$ in the unit of $M_p=1$. This requirement leads to a loose
lower bound on the curvaton mass $m\geq 2.8\times
10^{-37}f_{NL}^2/p^2$.

To summarize, the vacuum expectation value of curvaton is sensitive
to the physics in the very early universe. In order to get a
red-tilted and large non-Gaussian primordial power spectrum in
curvaton model we need to choose a suitable initial condition for
inflation. As we know, one of the advantage of inflation is that its
observational consequences are independent on the initial condition,
which make the predictions of inflation strong. Unfortunately the
curvaton model we discussed in this note loses this nice point. We
need to figure out a more reasonable model with a red-tilted and
large non-Gaussian primordial power spectrum in the future.

\vspace{.5cm}

\noindent {\bf Acknowledgments}

We would like to thank P. Chingangbam and M. Sasaki for useful
discussions.

\end{document}